\documentclass[12pt]{article}
\begin{document}
\noindent
{\Large \bf Statistical mechanical 
derivation of the
second law of thermodynamics}

\bigskip
\noindent
Hal Tasaki\footnote{
Department of Physics,
Gakushuin University,
Mejiro, Toshima-ku, Tokyo 171,
JAPAN

electronic address: 
hal.tasaki@gakushuin.ac.jp
}
%%%%%%%%%%%%%%%%%%%%%%%%%%%%%
\begin{abstract}
In a macroscopic (quantum or classical) Hamiltonian
system,
we prove the second law of thermodynamics 
in the forms of the minimum work principle and the
law of entropy increase, 
under the assumption that the initial state is
described by a general equilibrium distribution.
Therefore the second law is a logical necessity
once we accept equilibrium statistical mechanics.
\end{abstract}

\noindent
{\bf Note added} (December, 2000):
We learned that 
the main observation in the present note 
(the inequality (8) in the case
\( H=H' \))
is contained in 
A. Lenard, J. Stat. Phys.
{\bf 19}, 575 (1978).
(See also
W. Pusz and S. L. Woronowicz,
Commun. Math. Phys. {\bf 58}, 273 (1978),
W. Thirring,
{\em Quantum Mechanics of Large
Systems}\/
(Springer, 1983).)
We note, however, that Lenard does not
have the view point of micro-macro
separation (which we believe necessary 
for thermodynamic discussions)
and do not discuss
results like our
Theorem 2.
We do not publish the present note as an
original paper, but try to discuss these
(and other related) points in future
publications.

\bigskip
\bigskip
To understand macroscopic irreversibility 
from microscopic 
mechanics is one of the unsolved fundamental
problems in physics.
One may roughly classify the problem of irreversibility
into that of ``equilibration'' and 
of ``operational irreversibility.''
The former aims at justifying equilibrium statistical
mechanics, 
and has a rich (but not yet satisfactory)
history which goes back to Boltzmann 
\cite{note:history}.
The latter problem of 
``operational irreversibility''
deals with a microscopic interpretation of the
second law of thermodynamics.
Briefly speaking,
(a version of) the second law
is a statement about the fundamental limitation
on the possibility of adiabatic operation bringing
one equilibrium state to another \cite{LY}.
Recall that, although the initial and final states
are assumed to be in perfect equilibriums,
the process connecting the two can go far away from
equilibrium.

Although these two problems about irreversibility are
intimately connected,
we here concentrate only on the second problem of
``operational irreversibility'' and give a
solution to it.
More precisely, we assume that the initial 
sate of a macroscopic
(quantum or classical) Hamiltonian 
system is described by a general
equilibrium distribution, and rigorously 
derive the second law
of thermodynamics in the forms of the 
minimum work principle
and the law of entropy increase \cite{MWP}.
This establishes that {\em 
the second law is a logical 
necessity once we accept 
equilibrium statistical
mechanics and 
Hamiltonian mechanics}\/.

\paragraph*{Setup and basic inequality:}
We examine an adiabatic 
process in thermodynamics,
where an external (classical) 
agent performs an operation
on a thermally isolated system.
This may be modelled 
by a quantum mechanical system 
with an \( N \)-dimensional Hilbert space
whose time evolution is determined by
a time dependent
Hamiltonian \( H(t) \).
(One imagines that 
the agent changes some of the parameters in the
Hamiltonian.)
The operation  
(and the period of equilibration
following it) takes place in the time
interval 
\( t_{\rm init}\le t\le t_{\rm fin} \),
and we write
\( H=H(t_{\rm init}) \) and 
\( H'=H(t_{\rm fin}) \).
We denote by \( U \) the unitary time evolution 
operator for the whole operation \cite{note:U}.

For \( i=1,\ldots,N \), let 
\( |\varphi_{i}\rangle \)
and
\( |\varphi'_{i}\rangle \)
be the eigenstates of the  
Hamiltonians
\( H \) and \( H' \), respectively, with
the eigenvalues \( E_{i} \) and 
\( E'_{i} \).
The states are numbered so 
that 
\( E_{i}\le E_{i+1} \)
and
\( E'_{i}\le E'_{i+1} \).
We assume that the initial equilibrium
state
is described by a density matrix
\( \rho_{\rm init}=w(H)=
\sum_{i=1}^N |\varphi_{i}\rangle w(E_{i})
\langle\varphi_{i}| \),
where \( w(\cdot) \) is a function of energy.
We denote the initial expectation as
\( \langle\cdots\rangle_{\rm init}=
{\rm Tr}[(\cdots)\rho_{\rm init}] \),
and the final expectation
(corresponding to the unitary operator \( U \))
as
\( \langle\cdots\rangle_{U}=
{\rm Tr}[(\cdots)U\rho_{\rm init}U^{-1}] \).

We define a special unitary 
transformation \( \tilde{U} \)
by 
\( \tilde{U}|\varphi_{i}\rangle
=|\varphi'_{i}\rangle \) 
for
\( i=1,\ldots,N \).
The ``adiabatic theorem'' in quantum mechanics 
implies that
the unitary operator
\( \tilde{U} \) arises as the solution of 
the Schr\"{o}dinger
equation when the operation 
(i.e., the change of the 
Hamiltonian) is executed infinitely slowly.
Thus we may suppose that \( \tilde{U} \)
corresponds to a quasi-static operation in
thermodynamics \cite{note:qs}.

Our results are based on the following inequality
\cite{note:ineq}.

{\em Lemma}\/ --- 
Let \( w(E) \) be nonincreasing in \( E \),
and let \( F \) be an observable such that
\( F|\varphi'_{i}\rangle=F_{i}|\varphi'_{i}\rangle \)
with \( F_{i} \) nondecreasing in \( i \).
Then, for any \( U \),
\begin{equation}
	\langle F\rangle_{U}
	\ge
	\langle F\rangle_{\tilde{U}}.
	\label{eq:ff}
\end{equation}

{\em Proof:}\/
Observe that
\begin{eqnarray}
	\langle F\rangle_{U} 
	&=&
	{\rm Tr}[F\,U\,w(H)\,U^{-1}]
	\nonumber\\
	&=&
	\sum_{i,j=1}^N
	\langle\varphi'_{i}|
	F
	|\varphi'_{i}\rangle
	\,
	\langle\varphi'_{i}|
	U
	|\varphi_{j}\rangle
	\,
	\langle\varphi_{j}|
	w(H)
	|\varphi_{j}\rangle
	\,
	\langle\varphi_{j}|
	U^{-1}
	|\varphi'_{i}\rangle
	\nonumber\\
	&=&
	\sum_{i,j=1}^N
	F_{i}\,\alpha_{i,j}\,w_{j},
	\label{eq:fHpU}
\end{eqnarray}
where 
\( w_{j}=w(E_{j}) \),
and 
\( \alpha_{i,j}=
|\langle\varphi'_{i}|U|\varphi_{j}\rangle|^2 \).
Since \( \alpha_{i,j}=\delta_{i,j} \)
when \( U \) happens to be \( \tilde{U} \),
we find
\( \langle F\rangle_{\tilde{U}}
=\sum_{i=1}^N F_{i}\,w_{i} \).
From the unitarity, we have
\( 0\le\alpha_{i,j}\le1 \) and 
\( \sum_{i}\alpha_{i,j}=\sum_{j}\alpha_{i,j}=1 \).
By setting \( w_{N+1}=0 \),
we have for any \( A_{j} \) that 
\begin{equation}
	\sum_{j=1}^NA_{j}w_{j}
	=
	\sum_{j=1}^N(w_{j}-w_{j+1})
	\sum_{k=1}^jA_{k}.
	\label{eq:Aw}
\end{equation}
By using this we find
\begin{equation}
	\sum_{i,j=1}^N
	F_{i}\,\alpha_{i,j}\,w_{j}
	=\sum_{j=1}^N(w_{j}-w_{j+1})
	\sum_{i=1}^N\beta_{i}^{(j)}F_{i},
	\label{eq:faw}
\end{equation}
with 
\( \beta_{i}^{(j)}=\sum_{k=1}^j\alpha_{i,k} \),
which satisfies 
\( 0\le\beta_{i}^{(j)}\le1 \), and
\( \sum_{i=1}^N\beta_{i}^{(j)}=j \).
Since \( F_{i} \) is nondecreasing, we have
\begin{eqnarray}
	\sum_{i=1}^N\beta_{i}^{(j)}F_{i}
	&=&
	\sum_{i=1}^j\beta_{i}^{(j)}F_{i}
	+
	\sum_{i=j+1}^N\beta_{i}^{(j)}F_{i}
	\nonumber\\
	&\ge&
	\sum_{i=1}^j\beta_{i}^{(j)}F_{i}
	+
	(\sum_{i=j+1}^N\beta_{i}^{(j)})F_{j}
	\nonumber\\
	&=&
	\sum_{i=1}^j\beta_{i}^{(j)}F_{i}
	+
	(j-\sum_{i=1}^j\beta_{i}^{(j)})F_{j}
	\nonumber\\
	&\ge&
	\sum_{i=1}^j\beta_{i}^{(j)}F_{i}
	+
	\sum_{i=1}^j(1-\beta_{i}^{(j)})F_{i}
	=\sum_{i=1}^jF_{i}
	\label{eq:bf}
\end{eqnarray}
Substituting (\ref{eq:bf}) into (\ref{eq:faw})
and using (\ref{eq:Aw}), we get
\begin{equation}
	\sum_{i,j=1}^N
	F_{i}\,\alpha_{i,j}\,w_{j}
	\ge
	\sum_{j=1}^N(w_{j}-w_{j+1})\sum_{i=1}^jF_{i}
	=\sum_{i=1}^NF_{i}\,w_{i},
	\label{eq:faw2}
\end{equation}
which is the desired bound. 

\paragraph*{Physical quantities:}
We consider two physical quantities.
The work \cite{note:W} 
\( W(U) \) done by the agent to the system
during the operation (described by \( U \))
is 
\begin{equation}
	W(U)=\langle H'\rangle_{U}
	-\langle H\rangle_{\rm init}.
	\label{eq:WU}
\end{equation}
Note that this definition  follows uniquely from the 
energy conservation (of the system plus the agent).

We introduce the energy level operators
\( \Omega \) and \( \Omega' \) for the Hamiltonians
\( H \) and \( H' \), respectively,
by 
\( \Omega|\varphi_{j}\rangle=
j|\varphi_{j}\rangle \)
and
\( \Omega'|\varphi'_{j}\rangle=
j|\varphi'_{j}\rangle \).
Then we define the initial entropy as
\( S_{\rm init}=
\langle\log\Omega\rangle_{\rm init} \)
and the entropy after the operation as
\( S_{\rm fin}(U)=
\langle\log\Omega'\rangle_{U} \).
Note that 
\( \tilde{U}^{-1}\Omega'\tilde{U}=\Omega \),
and hence 
\( S_{\rm fin}(\tilde{U})=S_{\rm init} \).
Although there are various ways of 
defining entropies,
these are natural extensions of
the Boltzmann entropy \cite{note:SB}.
It is not hard to check that
\cite{Hal2}, for 
equilibrium states
of a macroscopic system,
the above definition agrees with other statistical 
mechanical entropies and hence with the
thermodynamic entropy.
It is when one considers non-equilibrium states,
that various statistical mechanical entropies
exhibit different behaviors.
See below.

\paragraph*{Second law for 
nonincreasing distributions:}
By setting \( F=H' \) or \( F=\log\Omega' \) ,
we can readily apply the 
inequality (\ref{eq:ff}) to get

{\em Theorem 1}\/ --- 
Suppose that \( w(E) \) 
(which determines the initial
distribution) is nonincreasing in \( E \).
Then we have the {\em minimum work principle}
\begin{equation}
	W(U)\ge W(\tilde{U}),
	\label{eq:WW}
\end{equation}
and
{\em the law of entropy increase}
\begin{equation}
	S_{\rm fin}(U)\ge S_{\rm fin}(\tilde{U})=S_{\rm init},
	\label{eq:SS}
\end{equation}
for any \( U \) which is consistent with the 
Hamiltonians \( H \) and \( H' \).

Among the examples of 
distributions with nonincreasing
\( w(E) \) are the {\em canonical distribution} with
\( w^{\rm can}_{\beta}(E)=
e^{-\beta E}/\sum_{i}e^{-\beta E_{i}} \),
and the {\em extended 
microcanonical ensemble} with
\( w^{\rm emc}_{\bar{E}}(E)=
\chi[E\le\bar{E}]/N(\bar{E}) \)
where 
\( N(\bar{E})=\max\{i|E_{i}\le\bar{E}\} \)
is the number of states.
(The characteristic function is defined by
\( \chi(\mbox{true})=1 \)
and \( \chi(\mbox{false})=0 \).)
In the extended microcanonical ensemble,
all the eigenstates with energies 
below \( \bar{E} \)
are given the equal weights.
Although this distribution may not be popular,
it is known to recover thermodynamics equally well
as the standard distributions.
(See, for example, \cite{Griffiths}.)
We remind the reader that
modern statistical physics 
relies fundamentally on the fact that
an equilibrium state of a 
{\em macroscopic}\/ system
can be {\em represented}\/ by various 
different statistical distributions, but 
the different descriptions provide us with 
{\em exactly}\/ the same thermodynamics
\cite{note:history}.

The inequality (\ref{eq:WW}) 
represents the {\em minimum work 
principle}\/ for adiabatic
processes, which says that 
the work required for a given
change of extensive parameters 
takes its minimum in the limit
of slow operation \cite{note:P}.
The minimum work principle is usually stated for
isothermal processes.
To get this version of the principle,
we consider a system which can be separated into 
a small (but still macroscopic) subsystem and
a big heat bath.
We decompose the Hamiltonian as
\( H=H_{\rm sub}+H_{\rm bath}+H_{\rm int} \)
and assume that the interaction 
\( H_{\rm int} \) 
between the subsystem and the bath is small 
in a certain sense.
When one only changes 
\( H_{\rm sub} \) during the operation,
we can show (by a standard estimate \cite{Hal2}) that
\( W(\tilde{U})\simeq 
F(\beta,H'_{\rm sub})-F(\beta,H_{\rm sub}) \)
where
\( F(\beta,H_{\rm sub})=
-\beta^{-1}\log{\rm Tr}[\exp(-\beta H_{\rm sub})] \)
is the free energy of the subsystem.
Then the inequality (\ref{eq:WW}) implies the
minimum work principle for an isothermal process.

The inequality (\ref{eq:SS}) represents
the {\em law of entropy increase}\/,
which says that the entropy never decreases
when an adiabatic process is possible
\cite{note:EP}.
This may be regarded as a realization
of the conventional wisdom \cite{Sinc} 
(which seems to be not rigorously founded)
that a ``coarse grained'' entropy 
can measure irreversibility \cite{note:GvsB}.
For the reader who worries about the arbitrariness
in the definition of entropy,
we remark that the law of entropy increase 
can also be derived from purely thermodynamic
considerations \cite{note:S}
by using the minimum work principle
(\ref{eq:WW}), which does not suffer from
any ambiguities in the definitions.

Therefore if we assume that the initial 
state is described
by the canonical or the extended microcanonical
(or other nonincreasing) distributions,
then the second law of thermodynamics
(i.e., the minimum work principle and the
law of entropy increase)
can be established as rigorous 
(and easily proved)
mathematical statements \cite{note:micro}.

\paragraph*{Microcanonical ensemble:}
We can also treat the standard microcanonical 
distribution with the weight
\( w^{\rm mc}_{\bar{E},\Delta E}(E)=
\chi[\bar{E}\le E\le\bar{E}+\Delta E]/
N(\bar{E},\Delta E) \)
where \( N(\bar{E},\Delta E) \)
is the number of levels such that
\( \bar{E}\le E_{i}\le\bar{E}+\Delta E \).
In this case, however, we do not have the inequality
(\ref{eq:ff}), but only have its weaker form
\cite{note:ff2}
\begin{equation}
	\langle F\rangle_{U}
	\ge
	\langle F\rangle_{\rm min}
	\equiv
	\frac{1}{N(\bar{E},\Delta E)}
	\sum_{i=1}^{N(\bar{E},\Delta E)}F_{i},
	\label{eq:ff2}
\end{equation}
where \( \langle\cdots\rangle_{\rm min} \) 
is the expectation
in the distribution where the  
\( N(\bar{E},\Delta E) \)
states with the lowest energies 
are given exactly the
same weights.
It is clear that 
\( \langle F\rangle_{\rm min} \) is in general
smaller than the desired
\( \langle F\rangle_{\tilde{U}} \).
But for a normal macroscopic systems and 
an extensive \( F \),
we can prove that 
\( \langle F\rangle_{\rm min} \) 
and
\( \langle F\rangle_{\tilde{U}} \)
are  essentially
identical with each other \cite{note:TDL}.
Consequently, we get

{\em Theorem 2}\/ --- 
Suppose that the model has an additional parameter
\( V \) (the volume), and we take a suitable
thermodynamic limit  \( V\to\infty \)
as is explained in \cite{note:TDL}.
Then we have the {\em minimum work principle}\/
and the {\em law of entropy increase}\/
in the forms
\begin{equation}
	\liminf_{V\to\infty}
	\{W(U)-W(\tilde{U})\}\ge0,
	\label{eq:Wmc}
\end{equation}	
and
\begin{equation}
	\liminf_{V\to\infty}
	\{S_{\rm fin}(U)-S_{\rm init}\}\ge0,
	\label{eq:Smc}
\end{equation}
for any \( U \) consistent with the Hamiltonians
\( H \) and \( H' \).

The discussions that follow Theorem 1 apply
equally here.

\paragraph*{Classical systems:}
All of the above results can be proved for
classical Hamiltonian systems as well.
Since the extension is straightforward
\cite{Hal2},
let us discuss only the basic ideas.

Consider exactly the same situation as in
quantum systems.
We denote a phase space point by \( \Gamma \),
and let \( H(\Gamma) \) and \( H'(\Gamma) \)
be the initial and the final 
Hamiltonians, respectively.
We assume that the initial state is sampled from
a probability distribution 
\( d\mu_{\rm init}=w(H(\Gamma))\,d\Gamma \) where
\( w(E) \) is a function 
of energy and \( d\Gamma \)
is the Lebesgue measure.
The time evolution map for the whole operation is
denoted as \( \tau \),
and the map corresponding to 
infinitely slow operation
is denoted as \( \tilde{\tau} \).
The relevant expectation values of a function
\( F(\tau \)) are
\( \langle F\rangle_{\rm init}=
\int d\mu_{\rm init}(\Gamma)\,F(\Gamma) \)
and
\( \langle F\rangle_{\tau}=
\int d\mu_{\rm init}(\Gamma)\,
F(\tau(\Gamma)) \).

In order to make use of the 
argument for quantum systems,
we decompose the phase space into a disjoint union 
\( \bigcup_{i=1}^\infty V_{i} \) with
\( V_{i}=\{\Gamma\,|
\,E_{i}\le H(\Gamma)<E_{i+1}\} \).
Here the sequence \( E_{1}, E_{2},\ldots \)
is chosen so that \( |V_{i}|=v \) 
with a constant \( v>0 \) for all \( i \).
(\( |V| \) is the 
Lebesgue volume of a set \( V \).)
We also construct a similar decomposition
\( \bigcup_{i=1}^\infty V'_{i} \) 
for \( H'(\Gamma) \) using the same volume \( v \).
Then, for a given \( \tau \), we define
\( \alpha_{i,j}=\delta^{-1}|
V'_{i}\cap\tau(V_{j})| \),
which is the ratio of the pase space point in
\( V_{j} \) which flow into \( V'_{i} \).
By using the Liouville theorem (i.e., the conservation
of the phase space volume),
we see that \( \alpha_{i,j}\ge0 \),
and \( \sum_{i}\alpha_{i,j}
=\sum_{j}\alpha_{i,j}=1 \)
exactly as in quantum cases.
The adiabatic theorem
again implies that 
\( \alpha_{i,j}=\delta_{i,j} \)
for the map \( \tilde{\tau} \).

Assume that the function \( F(\Gamma) \) 
depends only on the energy \( H'(\Gamma) \).
Since
\( H(\Gamma)=E_{j}+O(v) \) 
for \( \Gamma\in V_{j} \),
\( H'(\Gamma)=E'_{i}+O(v) \) 
for \( \Gamma\in V'_{i} \),
and 
\( \sum_{i,j}\chi[\Gamma\in V_{j},
\tau(\Gamma)\in V'_{i}]=1 \),
we find that
\begin{eqnarray}
	\langle F\rangle_{\tau}
	&=&
	\sum_{i,j}\int d\Gamma\,F(\Gamma)\,
	\chi[\Gamma\in V_{j},
	\tau(\Gamma)\in V'_{i}]
	\,
	w(H(\Gamma))
	\nonumber\\
	&\simeq&v\sum_{i,j}F(E'_{i})\,
	\alpha_{i,j}\,w(E_{j}).
	\label{eq:Fclassical}
\end{eqnarray}
Note that the final expression becomes exact as 
\( v\to0 \).
Observing that this representation is exactly
the same as the quantum one,
the rest of the proof proceeds automatically.

\paragraph*{Discussions:}
We have presented general theorems which
establish the validity of the second law of
thermodynamics (in the forms 
of minimum work principle
and the law of entropy increase)
under the assumption that the initial state is 
described by a certain equilibrium distribution.
This means that the ``order theoretical''
nature \cite{LY} of the second law is
already inherent in 
equilibrium statistical mechanics 
and Hamiltonian mechanics.

A major and essential limitation to
the present
approach is that the final 
state with the density
matrix 
\( \rho_{\rm fin}=U\rho_{\rm init}U^{-1} \) 
in general does not correspond to an
{\em exact}\/ equilibrium distribution
(while we expect it to describe an equilibrium
state in a {\em macroscopic}\/ sense).
This means that our derivation of the second law
is not useful if we perform yet another operation
after the first one 
(where \( \rho_{\rm fin} \) determines
the initial state).
Given the fact that 
thermodynamics has been verified
in experiments repeated over and over in the
history,
we have to say that this limitation is quite
serious.

A solution to this problem is simply to replace 
\( \rho_{\rm fin} \) with an exact equilibrium 
density matrix
\( \tilde{\rho}_{\rm fin} \),
where the latter is chosen so that the replacement 
produces no macroscopically observable effects.
Then we can apply the argument of the present paper
to derive the second law.
Of course the main remaining issue here is to 
justify \cite{note:eq} the
replacement of 
\( \rho_{\rm fin} \) by \( \tilde{\rho}_{\rm fin} \),
which is a part of the  problem of 
``equilibration.''

%%%%%%%%%%%%%%%%%%%%%%%%%%%%%%%%%%%%%%%%%%%
%%%%%%%%%%%%%%%%%%%%%%%%%%%%%%%%%%%%%%%%%%%
\bigskip
It is a pleasure to thank
Shin-ichi Sasa
for various useful discussions and suggestions
which have been essential for the present work,
Elliott Lieb for valuable discussions 
and comments,
and 
Yoshi Oono,
Joel Lebowitz, 
and
Ken Sekimoto
for 
useful discussions on related topics.
%%%%%%%%%%%%%%%%%%%%%%%%%%%%%%%%%%%%%%%%%%%
%%%%%%%%%%%%%%%%%%%%%%%%%%%%%%%%%%%%%%%%%%%

%%%%%%%%%%%%%%%%%%%%%%%%%%%%%%%%%%%%%%%%%%%
%%%%%%%%%%%%%%%%%%%%%%%%%%%%%%%%%%%%%%%%%%%
%%%%%%%%%%%%%%%%%%%%%%%%%%%%%%%%%%%%%%%%%%%

\begin{thebibliography}{10}

\bibitem{note:history}
Since the literature  is
so vast, we only refer to a recent monograph;
G. Gallavotti,
{\em Statistical Mechanics: A Short Treatise}\/,
(Springer, 1999).

\bibitem{LY}
E. H. Lieb and J. Yngvason,
Phys. Rep. {\bf 310}, 1 (1999);
Physics Today, April 2000, p.~32.


\bibitem{MWP}
For classical Hamiltonian systems with
canonical distribution,
some versions of 
the second law has been proved in
C. Jarzynski, Phys. Rev. Lett. 
{\bf 78}, 2690 (1997);
J. Stat. Phys. {\bf 96}, 415 (1999).
Extensions to quantum systems
is not hard.

\bibitem{note:N}
Since we are proving only inequalities,
it is trivial to take the limit \( N\to\infty \).
One can treat, for example, systems of particles
in a continuum space 
by first introducing artificial ``cutoffs''
to make the Hilbert space finite dimensional,
and carefully taking necessary limits
after the proofs.

\bibitem{note:U}
Let \( U(t) \) be the solution of the Schr\"{o}dinger 
equation
\( i\partial U(t)/\partial t=H(t)U(t) \)
with \( U(t_{\rm init})={\bf 1} \).
Then \( U=U(t_{\rm fin}) \).



\bibitem{note:qs}
This conclusion is not trivial as it appears.
In a usual macroscopic system, the adiabatic theorem
requires an operation to be done so slowly that it
can never be (thermodynamically) realistic.
Nevertheless we expect the theorem to provide us
with information of 
thermodynamic quasi-static operations 
unless 
the operation contains nontrivial steps such as
thermally decoupling a part of the system.

\bibitem{note:ineq}
The following inequality and its implication to the
second law was first discussed in
H. Tasaki, unpublished note (cond-mat/0005128),
where we focused on partial derivation of the second 
law directly from quantum dynamics
in systems satisfying certain conditions.
(See H. Tasaki, Phys. Rev. Lett. {\bf 80}, 1373 (1998).)
We then realized that the argument can be used more
effectively in the present context.
See \cite{Hal2} for a unified treatment.

\bibitem{Hal2}
H. Tasaki, to be published.

\bibitem{note:W}
Note that \( W(U) \) is the 
{\em expectation value}\/ of the energy difference.
The actual work does fluctuate around \( W(U) \),
but the fluctuation must be negligible in
nonpathological macroscopic states.

\bibitem{note:SB}
The Boltzmann entropy is defined as
\( S_{\rm B}=\log W \)
(we set the Boltzmann constant to be unity),
where \( W \) is the number of microstates
with the same prescribed macroscopic characters.
Now given a generic eigenstate with energy \( E \)
of a macroscopic quantum system,
we can choose \( W \) to be either
the density of states at \( E \) or the number
of eigenstates with energies below \( E \).
Both give the same result in the thermodynamic
limit, and our definition corresponds to
the latter choice.
The dependence of our entropy on 
Hamiltonian corresponds to a 
macroscopic characterization in terms
of energy variables.



\bibitem{Griffiths}
R. B. Griffiths,
J. Math. Phys.~{\bf 6}, 1447 (1965);
E. H. Lieb and J. Lebowitz,
Adv. Math. {\bf 9}, 316 (1972).


\bibitem{note:P}
For the special case with \( H'=H \),
we have \( \tilde{U}={\bf 1} \) 
and hence \( W(\tilde{U})=0 \).
Then (\ref{eq:WW}) becomes 
\( W(U)\ge0 \),
which means that the work done 
during an adiabatic cycle
cannot be negative.
This is  the Planck's principle \cite{LY}.

\bibitem{note:EP}
The {\em entropy principle}\/
states that an adiabatic process is possible
{\em when and only when}\/ the entropy does
not decrease \cite{LY}.
See \cite{Hal2} for the remaining
``when'' part.

\bibitem{Sinc}
See, for examples,
D. N. Zubarev,
{\em Nonequilibrium Statistical Thermodynamics}\/
(Plenum, New York, 1974),
Section 8.4;
J. Rau and B. M\"{u}ller, Phys. Rep.
{\bf 272}, 1 (1996),
Section A.3.1.



\bibitem{note:GvsB}
Recall that the von Neuman and the Gibbs
entropies are time independent.
But we believe that this fact independence
no significance in thermodynamics
since these entropies coincide
with the thermodynamic entropy only
when the state corresponds to an
{\em exact}\/ equilibrium distribution.
See, for example, 
J. Lebowitz,
Physics Today, September 1993, p.~32.


\bibitem{note:S}
Denote by \( X \) and \( X' \) the
sets of extensive variables corresponding to
\( H \) and \( H' \), respectively.
We assume (as usual) that a 
thermodynamic equilibrium
state is fully determined by specifying its 
energy and extensive variables.
Let \( E=\langle H\rangle_{\rm init} \),
\( E'=\langle H'\rangle_{U} \),
and \( \bar{E}'=\langle H'\rangle_{\tilde{U}} \).
The thermodynamic entropy should satisfy
\( S_{\rm TD}(E,X)=S_{\rm TD}(\bar{E}',X') \)
because the two states are connected by 
a quasi-static adiabatic process.
We also have
\( S_{\rm TD}(E',X')\ge S_{\rm TD}(\bar{E}',X') \)
since \( E'\ge\bar{E}' \) from (\ref{eq:WW}),
and the entropy is increasing in energy.
Then we get the desired inequality
\( S_{\rm TD}(E',X')\ge S_{\rm TD}(E,X) \).

\bibitem{note:micro}
Note that the system need not be macroscopic.
This fact may find applications in the discussions
of ``small'' systems.  See
K. Sato, K. Sekimoto, T. Hondou, and F. Takagi,
preprint (cond-mat/0008393),
and
H. Tasaki, unpublished note (cond-mat/0008420),
where we already used the result of the
present note.


\bibitem{note:ff2}
The inequality is almost trivial, but let us preset
a proof.
We abbreviate \( N(\bar{E}) \) and \( N(\bar{E},\delta E) \) 
as \( N_{0} \) and \( \bar{N} \), respectively.
Then
\( \langle F\rangle_{U}
=\bar{N}^{-1}\sum_{j=N_{0}+1}^{N_{0}+\bar{N}}
\sum_{i=1}^NF_{i}\,\alpha_{i,j}
=\bar{N}^{-1}\sum_{j=1}^{\bar{N}}
\sum_{i=1}^NF_{i}\,\tilde{\alpha}_{i,j},
\)
where
we defined
\( \tilde{\alpha}_{i,j}=\alpha_{i,j+N_{0}} \)
for \( j=1,\ldots,\bar{N} \).
We have
\( \tilde{\alpha}_{i,j}\ge0 \),
\( \sum_{i=1}^N\tilde{\alpha}_{i,j}
=1 \),
and
\( \sum_{j=1}^{\bar{N}}\tilde{\alpha}_{i,j}\le1 \).
Then (\ref{eq:ff2}) can be proved 
exactly as (\ref{eq:faw2}).

\bibitem{note:TDL}
{\em Sketch of the assumptions and proof:}\/
We assume that there is an additional parameter 
\( V \) (the volume), and the numbers of states
\( N(E) \) and \( N'(E') \) for \( H \) and \( H' \),
respectively,
behave as 
\( N(\varepsilon V)\simeq e^{V\sigma(\varepsilon)} \)
and
\( N'(\varepsilon' V)\simeq e^{V\sigma'(\varepsilon')} \)
as \( V\to\infty \) with
continuous increasing functions (entropies)
\( \sigma(\varepsilon) \) and 
\( \sigma'(\varepsilon') \).
We choose the relevant energies as
\( \bar{E}=\bar{\varepsilon}V \)
and
\( \Delta E=\alpha V^{\delta} \)
with constants
\( \bar{\varepsilon} \), \( \alpha>0 \),
and \( 0<\delta<1 \).
The initial state 
\( \langle\cdots\rangle_{\rm init} \)
is the microcanonical distribution with the energy
range from \( \bar{E} \) to  
\( \bar{E}+\Delta E \),
the final state 
\( \langle\cdots\rangle_{\tilde{U}} \)
is that with the range 
 \( \bar{E}' \) to  \( \bar{E}'+\Delta E' \),
 and  \( \langle\cdots\rangle_{\rm min} \)
 is that with the range
 \( 0 \) to \( \tilde{E}' \) 
(where the lowest energy is set to zero).
Here \( \bar{E}' \), \( \Delta E' \),
and  \( \tilde{E}' \) are determined by
\( N'(\bar{E}')=N(\bar{E}) \),
\( N'(\bar{E}'+\Delta E')=N(\bar{E}+\Delta E) \),
and
\( N'(\tilde{E}')=N(\bar{E},\Delta E)
=N(\bar{E}+\Delta E)-N(\bar{E})+1 \).
By using the above assumptions,
we find that
\( \bar{E}'/V \), \( (\bar{E}'+ \Delta E')/V \),
and \( \tilde{E}'/V \) converge to a single value 
\( \bar{\varepsilon}' \)
as \( V\to\infty \).
Then if \( F(E')=Vf(E'/V) \),
the standard estimate (based on the fact that 
\( N'(E') \) is a rapidly increasing function)
implies
\( \langle f(H'/V)\rangle_{\tilde{U}}
\simeq f(\bar{\varepsilon}') \)
and
\( \langle f(H'/V)\rangle_{\rm min}
\simeq f(\bar{\varepsilon}') \)
as \( V\to\infty \).

\bibitem{note:eq}
Note that such a justification can only be
possible for {\em generic}\/ operations,
since the entropy does decrease for the
very special operation corresponding
to \( U^{-1} \).


\end{thebibliography}
\end{document}